\documentclass{optica-article}

\journal{opticajournal} 

\articletype{Research Article}

\usepackage{physics}
\begin{document}

\title{Manipulating the Quasi--Normal Modes of Radially Symmetric Resonators}

\author{James R Capers,\authormark{1,*} Dean A Patient,\authormark{1} and Simon A R Horsley\authormark{1}}

\address{\authormark{1}Department of Physics and Astronomy, University of Exeter, Stocker Road, Exeter, EX4 4QL, United Kingdom}

\email{\authormark{*}j.capers@exeter.ac.uk} 


\begin{abstract*} 
    We derive two methods for simultaneously controlling the resonance frequency, linewidth and multipolar nature of the resonances of radially symmetric structures.  
    Firstly, we formulate an eigenvalue problem for a global shift in the permittivity of the structure to place a resonance at a particular complex frequency.  
    Next, we employ quasi-normal mode perturbation theory to design radially graded structures with resonances at desired frequencies.  
\end{abstract*}

\section{Introduction}

Resonances are key to a strong electromagnetic response of photonic components.
Over the last two decades resonant interaction has been employed in metamaterials \cite{Munk2000, Staude2013, Hentschel2023}, nano--antennas \cite{Novotny2011, Bharadwaj2009} and super--scatterers \cite{Powell2020, Finlayson2023} to almost arbitrarily manipulate electromagnetic radiation.
Recently there has been much interest in `Mietronics' \cite{Won2019, Koshelev2021}.
Work in this field has focused on using higher order multipolar modes to achieve greater control of electromagnetic radiation including directional scattering \cite{Fu2013}, field confinement  \cite{Miroshnichenko2015} and frequency selective imaging \cite{Tittl2018}.

Engineering the resonant frequency of any particular multipolar response is crucial for many applications including efficient absorbers \cite{Aydin2011, Landy2008} and thermal emitters for gas sensing \cite{Lochbaum2018}.
Although the theory of Mie resonances is over 100 years old \cite{Mie1908}, the problem of designing the spectral response of multipolar resonances remains open.

Many approaches rely on a brute force sweep of the design parameters using memory intensive numerical simulations \cite{Grigoriev2013, Wu2020} to move resonances, which can be numerically demanding and offers little insight into the final structure.
More advanced methods based upon machine learning \cite{Real2022, Li2023} have recently been used to manipulate both electric and magnetic multipole moments, however a large amount of training data must be generated.
In addition to this, semi--analytic techniques based upon complex analysis \cite{Binkowski2022} have been developed to enhance the Q--factor of resonances.
This exploits the insight provided by the quasi--normal mode framework.

First developed by Gamow to describe alpha decay \cite{Gamow1928}, quasi--normal modes have been employed extensively to model electromagnetic systems in terms of their complex frequency resonances \cite{Kristensen2020, Ge2014, Lalanne2019}.
Crucially, for well separated quasi normal mode resonances, the real part of the quasi--normal mode frequency describes the resonance frequency, whereas the imaginary part encodes the spectral width of the resonance.
The ability to manipulate the exact location of quasi--normal modes in the complex plane therefore allows one to control the location and width of the resonance simultaneously.
Inverse design techniques to manipulate spatial electromagnetic responses are well developed \cite{Molesky2018, Lalau-Keraly2013, Capers2021, Capers2022a}, however few systematic techniques exist to manipulate the spectral response of resonators.

In this work, we develop two techniques to control the spectral location, width, and multipolar nature of the resonances of radially symmetric resonators.
Building on previous work that developed a design theory for 1D graded structures \cite{Capers2022}, we propose a technique to calculate the complex shift one should apply to the permittivity of a resonator to place a particular multipolar resonance at a particular complex frequency.
We also employ quasi--normal mode perturbation theory to design graded structures with particular resonances at particular frequencies.
Both techniques are simple, easily extensible and numerically efficient.

\section{Finding the Quasi--Normal Modes of Radially Symmetric Resonators \label{sec:finding_qnm}}

In this section, we briefly review how one can formulate a finite difference scheme to find the quasi--normal modes of a resonator with a radially graded permittivity.
Our key results are how these permittivity profiles can be designed, however it will be necessary to find the quasi--normal modes throughout.

The scalar Helmholtz equation governs a wide variety of physical phenomena.
For pressure acoustics and a single polarisation of the electromagnetic field in 2D it is exact, and is commonly used an approximation that neglects polarisation in 3D electromagnetism. 
For simplicity we will work with the scalar Helmholtz equation throughout.
We write the scalar Helmholtz equation as 
\begin{equation}
    \left[ \nabla^2 + k^2 \varepsilon (r) \right] \psi (\boldsymbol{r}) = 0 ,
    \label{eq:hh}
\end{equation}
describing a single polarisation of the electromagnetic field $\psi (\boldsymbol{r})$ in a material with spatially varying permittivity $\varepsilon(\boldsymbol{r})$ and wavenumber $k$.
Our aim is to find, for a given permittivity profile, the complex eigen--frequencies supported by the structure.
Simply re--arranging the Helmholtz equation (\ref{eq:hh}), this problem can be formulated as an eigenvalue problem 
\begin{equation}
    -\frac{1}{\varepsilon} \nabla^2 \psi = k^2 \psi .
    \label{eq:hh_ev}
\end{equation}
For this eigenvalue problem to give the complex quasi--normal mode frequencies, one must impose the out--going wave boundary condition.
This makes the Laplace operator depend upon the eigenvalue $k$ \cite{Kristensen2020, Capers2022}, meaning that the solution of Eqn. (\ref{eq:hh_ev}) is no longer a straightforward eigenvalue problem \cite{Lalanne2019}.
Here, we formulate how this eigenvalue problem can be solved for radially symmetric systems.

For a given radially symmetric permittivity profile $\varepsilon (\boldsymbol{r}) = \varepsilon (r)$ our aim is to find the locations of the quasi--normal modes supported by the resonator.
As we are treating radially symmetric resonators, we write the Laplacian in Eqn. (\ref{eq:hh}) in cylindrical or spherical coordinates 
\begin{equation}
    \nabla^2 \psi = 
    \begin{cases}
        \pdv[2]{\psi}{r} + \frac{1}{r} \pdv{\psi}{r} + \frac{1}{r^2} \pdv[2]{\psi}{\theta} \ & {\rm 2D} \\
        \pdv[2]{\psi}{r} + \frac{2}{r} \pdv{\psi}{r} + \frac{1}{r^2 \sin^2 \theta} \pdv{\psi}{\theta} \left( \sin \theta \pdv{\psi}{\theta} \right) + \frac{1}{r^2 \sin^2 \theta} \pdv[2]{\psi}{\phi}\ & {\rm 3D}
    \end{cases}
\end{equation}
depending on the number of dimensions we choose to work in.
While the calculations are very similar in 2D and 3D, for clarity we proceed with the 3D example providing the 2D calculation in the supplementary material.
Separating the angular and radial variables, we write $\psi (\boldsymbol{r}) = \psi (r) Y_{\ell}^m (\theta, \phi)$, where $Y_{\ell}^m (\theta, \phi)$ are the spherical harmonics.
Substituting this into Eqn. (\ref{eq:hh}) turns the Helmholtz equation into
\begin{equation}
    \pdv[2]{\psi}{r} + \frac{2}{r} \pdv{\psi}{r} - \frac{\ell (\ell+1)}{r^2} \psi (r) + k^2 \varepsilon (r) \psi (r) = 0.
\end{equation}
We now remove the first derivative term with the substitution $\psi (r) = \chi (r)/r$, to get
\begin{equation}
    \pdv[2]{\chi}{r} - \frac{\ell(\ell+1)}{r^2} \chi + k^2 \varepsilon (r) \chi = 0.
\end{equation}
This has now brought the equation we must solve into the form of the 1D Helmholtz equation, with an additional term related to the angular momentum $\ell$.
The boundary condition for an outgoing wave that must be imposed is
\begin{align}
    \psi (r \rightarrow \infty) &= h_\ell^{(1)} (kr) , \\
    \chi (r \rightarrow \infty) &= r h_\ell^{(1)} (kr) ,
\end{align}
where $h_\ell^{(1)} (kr)$ is the spherical Hankel function of the first kind.
Asymptotically, as $kr \rightarrow \infty$ the spherical Hankel function goes as $h_\ell^{(1)} (kr) \rightarrow i^{-(n+1)} e^{ikr} / (kr)$ \cite{dlmf}.
This means that the boundary condition upon the derivative of the field can be written as
\begin{align}
    \pdv{\chi}{r} &= h_\ell^{(1)} (kr) \left[ 1 + (kr) \frac{h_\ell^{(1)'} (kr)}{h_\ell^{(1)} (kr)}\right] \\
    &= \chi (r) \left[ \frac{1}{r} + k \frac{h_\ell^{(1)'} (kr)}{h_\ell^{(1)} (kr)}\right] \\ 
    &= \chi \gamma (k) .
    \label{eq:bc}
\end{align}
To impose this boundary condition, we shall need to modify the finite difference matrix that will numerically represent the Laplacian operator.
We therefore write the boundary condition given by Eqn. (\ref{eq:bc}) in forwards finite difference form
\begin{equation}
    \chi_{n+1}  = \chi_n (1+\Delta r \gamma (k)) .
\end{equation}
At this point, the boundary condition corresponding to quasi--normal modes can be imposed on the Laplacian.
However, to find the modes we must remove the dependence upon $k$ from the finite difference Laplacian matrix.
This can be achieved by linearising $\gamma (k)$ around a particular frequency $k_\star$, giving
\begin{equation}
    \gamma(k) = \gamma (k_\star) + (k-k_\star) \partial_k \gamma (k_\star) = A + k B ,
\end{equation}
where $A = \gamma (k_\star) - k_\star \partial_k \gamma (k_\star)$ and $B = \partial_k \gamma (k_\star)$.
We can now formulate the eigenvalue problem posed by Eqn. (\ref{eq:hh_ev}) such that the eigenvalue $k$ only appears on the right--hand side.
Defining 
\begin{align}
    \mathcal{L} &= \frac{1}{\Delta r^2}
    \begin{pmatrix}
        -2 & 1 & 0 & \cdots & 0 \\
        1 & -2 & 1 & \cdots & 0 \\
        0 & 1 & -2 & \cdots  & 0 \\
        \vdots & \vdots & \vdots & \ddots  & \vdots \\
        0 & 0 & 0 & \cdots & (A \Delta r - 1)
    \end{pmatrix} ,
    &
    \mathcal{L}' &= \frac{1}{\Delta r^2}
    \begin{pmatrix}
        0 & 0 & 0 & \cdots & 0 \\
        0 & 0 & 0 & \cdots & 0 \\
        0 & 0 & 0 & \cdots & 0 \\
        \vdots & \vdots & \vdots & \ddots & \vdots \\
        0 & 0 & 0 & \cdots & B \Delta r 
    \end{pmatrix} .
\end{align}
we can split the Helmholtz equation into
\begin{equation}
    \left( \mathcal{L} + k \mathcal{L}'  - \frac{\ell(\ell+1)}{r^2} + k^2 \varepsilon (r) \right) \chi = 0 .
\end{equation}
This can then be formed into a quadratic eigenvalue problem \cite{Tisseur2001}
\begin{equation}
    \begin{pmatrix}
        \boldsymbol{0} & \boldsymbol{1} \\
        -(\mathcal{L} - \ell(\ell+1)/r^2) / \varepsilon(r) & -\mathcal{L'} / \varepsilon(r)
    \end{pmatrix}
    \begin{pmatrix}
        \chi \\
        k \chi
    \end{pmatrix}
    = k
    \begin{pmatrix}
        \chi \\
        k \chi
    \end{pmatrix} .
\end{equation}
For radially graded structures, this gives both the radial part of the field of the mode and the eigenfrequency $k$.

To verify this method, we compare it to other methods for finding quasi--normal modes.
For isotropic cylinders and spheres, Mie theory can be used to find the complex frequencies analytically.
Full--wave solvers such as COMSOL Multiphysics \cite{COMSOL} can be employed to find the complex eigenfrequencies of spatially varying structures.
Both COMSOL and the method we have outlined here require frequencies to search around, so one must already have an idea of roughly where the resonance is located.
\begin{figure}[h!]
    \centering
    \includegraphics[width=\linewidth]{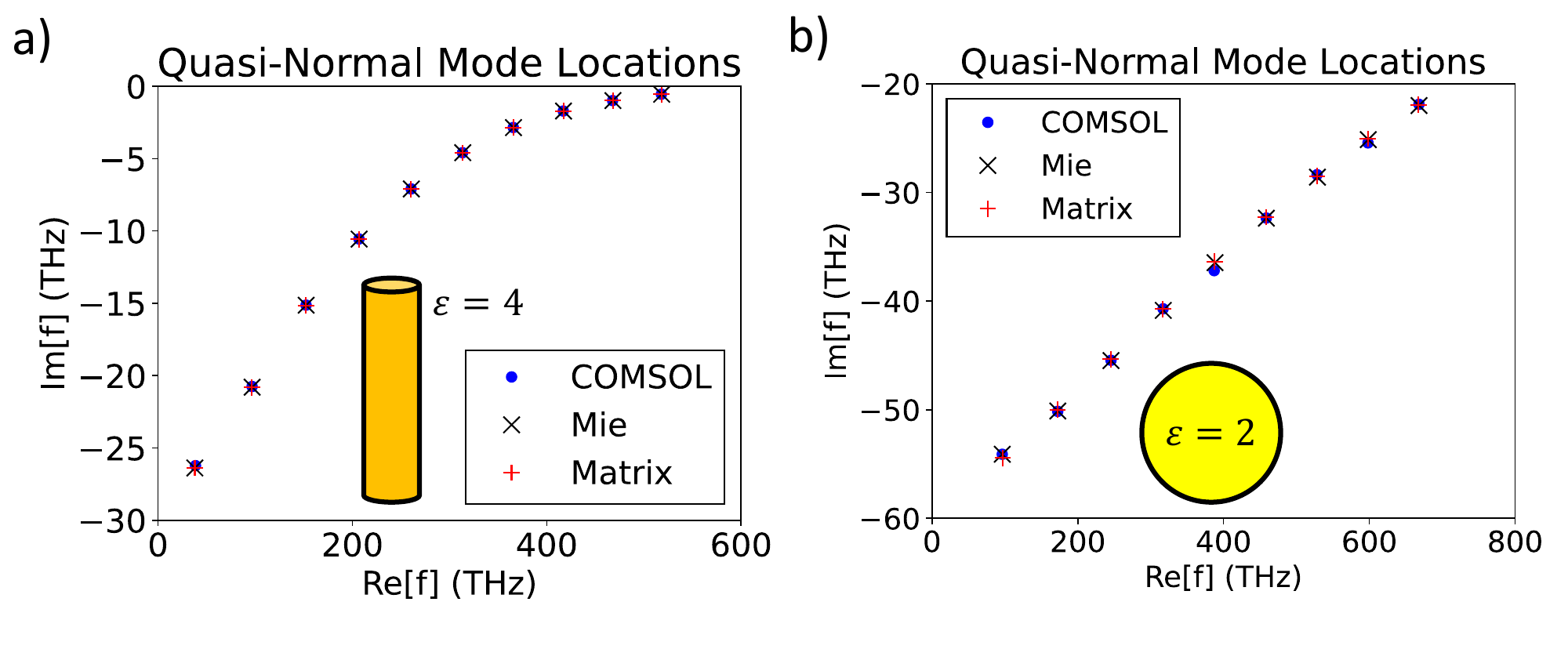}
    \caption{A comparison of different methods for finding the quasi--normal modes of isotropic cylinders or spheres.  a) shows the comparison for a cylinder, and b) for a sphere.
    Both the cylinder and the sphere have radius $a = 550$ nm.}
    \label{fig:comparison}
\end{figure}
Figure \ref{fig:comparison} shows the comparison of the three methods of finding the quasi--normal mode frequencies of isotropic cylinders and spheres.
We have chosen the cylinder and sphere to have radius $a = 550$ nm, with the cylinder having a permittivity of $\varepsilon = 4$ and the sphere $\varepsilon = 2$.
Details of the Mie theory are given in Supplementary Material.

With a method of finding both the complex frequency and mode profile of the quasi-normal modes of cylinders and spheres, we proceed to consider two methods for engineering the permittivity such that resonances are placed at particular complex frequencies.
This allows one to achieve a high level of control over the spectral response of a scatterer.

\section{Shifting Resonances By Finding `Eigen--Permittivities'}

Our first strategy for imposing a given quasi--normal mode frequency on a structure is calculating an overall shift to the permittivity to place a quasi--normal mode at a particular frequency.
To this end, we split the permittivity into a a spatially varying part $\varepsilon (r)$ and a background shift $\varepsilon_b$.
This allows us to rearrange the Helmholtz equation (\ref{eq:hh}) into an eigenvalue problem for the background permittivity
\begin{align}
    \left[ \nabla^2 + k^2 ( \varepsilon (r) + \varepsilon_b) \right] \phi &= 0 , \\
    -\frac{1}{k^2} \left[ \nabla^2 + k^2 \varepsilon (r) \right] = \varepsilon_b \phi . 
    \label{eq:ev}
\end{align}
To use this, one can choose a complex $k$ at which one wants the quasi--normal mode to occur, then solve the eigenvalue problem to find the corresponding (complex) permittivity shift $\varepsilon_b$.
Unlike previous results \cite{Capers2022}, for radially symmetric resonators the Laplacian contains information about the angular momentum of the mode that is placed at the desired $k$.
One can therefore control both the spectral properties and the multipolar nature of the resonance simultaneously.
It is important to note that in order to solve the eigenvalue problem posed by Eqn. (\ref{eq:ev}), the finite difference Laplacian operator must be modified to have the outgoing wave boundary condition at the edge of the resonator, as was discussed in Section \ref{sec:finding_qnm}.

\begin{figure}[h!]
    \centering
    \includegraphics[width=\linewidth]{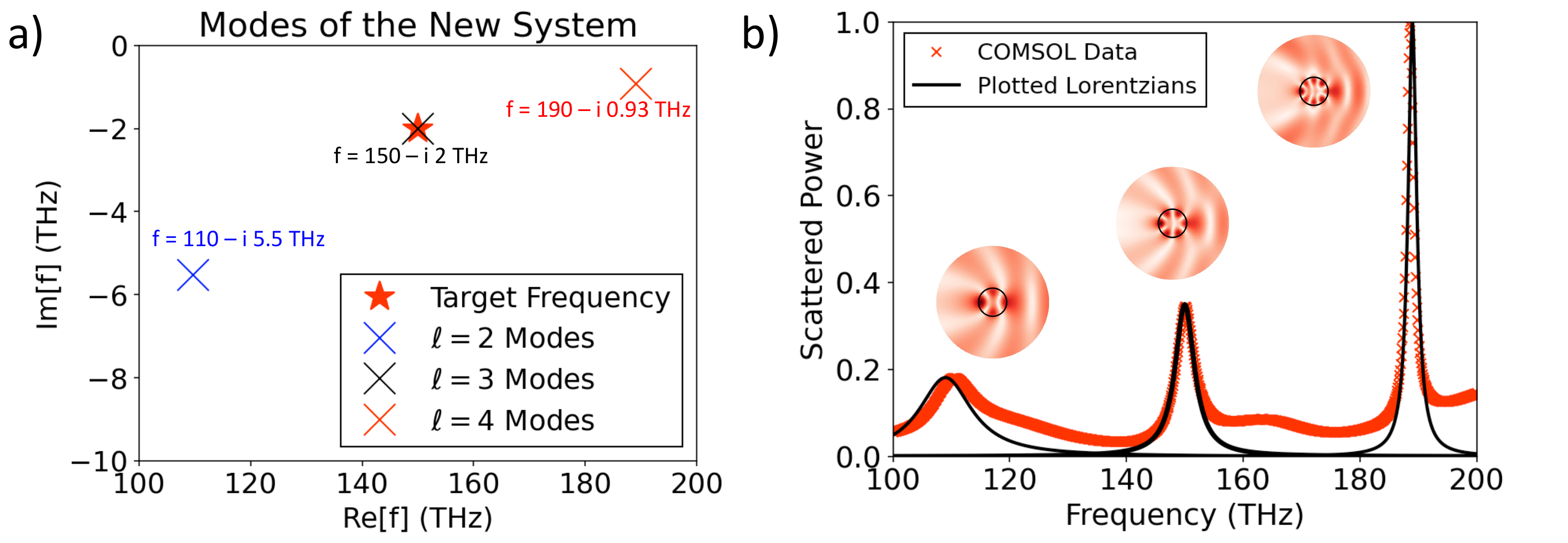}
    \caption{Using the eigen--permittivity method to place the $\ell=3$ mode of an isotropic cylinder.  
    We seek to move the $\ell=3$ mode to 150-i2 THz, requiring a background permittivity of $\varepsilon_b = 4.15+i0.007$.  
    a) The locations of the quasi--normal modes.
    b) Full--wave simulations verifying the scattering behaviour of the cylinder.  
    Peak locations and widths can be related directly to pole locations, and the fields $|\psi|$ shown inset verify the multipolar nature of the modes.  
    As the inset plots are of the field norm, the $\ell = 3$ lobe should exhibit 6 amplitude peaks.}
    \label{fig:cylinder_l3}
\end{figure}

An example of this procedure for a cylinder is shown in Figure \ref{fig:cylinder_l3}.
Starting from an isotropic cylinder of radius $a = 550$ nm and permittivity $\varepsilon = 4$, the $\ell = 3$ resonance is initially at $f \sim 200 - i 10$ THz.
We seek to move this to $f = 150 - i 2$ THz, corresponding to a lower resonance frequency and smaller linewidth, giving a background shift of $\varepsilon_b = 4.15 + i 0.007$.
Applying this, we find the $\ell=3$ mode at the desired frequency.
To verify this, full--wave simulations are employed to calculate the scattered energy from the cylinder.
It is well--known that the spectral response of a resonator with several well spaced quasi--normal modes can be approximated as a series of Lorentzians \cite{Kristensen2020}, with the peak locations corresponding to the real part of the quasi--normal modes and the widths corresponding to the imaginary parts.
Thus, to verify that moving the resonances has had the desired effect on the scattering from the resonator, we plot Lorentzians of the form
\begin{equation}
    L (f, f_0, \Gamma) = A \frac{\Gamma}{(f-f_0)^2 + \Gamma^2}
\end{equation}
over the scattered power data.
Throughout, the central frequency $f_0$ is taken from the real part of the mode with the width $\Gamma$ corresponding to the expected imaginary part and $A$ is the amplitude scale.
At $f = 150$ THz there is a peak corresponding to the $\ell=3$ quasi--normal mode, with a width corresponding to the imaginary part of the quasi--normal mode frequency.
The behaviour between the main three resonances can also be understood using the quasi--normal mode framework.
For example, at $f = 120 - i 12$ THz there is a $\ell = 0$ mode, which contributes the broad bump to the spectrum at around 120 THz.
Examining the field profile of the resonator, shown inset, we observe that the multipolar nature of the mode is as expected.
Other peaks in the spectrum are explained by the presence of the $\ell=2$ and $\ell=4$ modes in the vicinity of the mode we have moved.

\begin{figure}[h!]
    \centering
    \includegraphics[width=\linewidth]{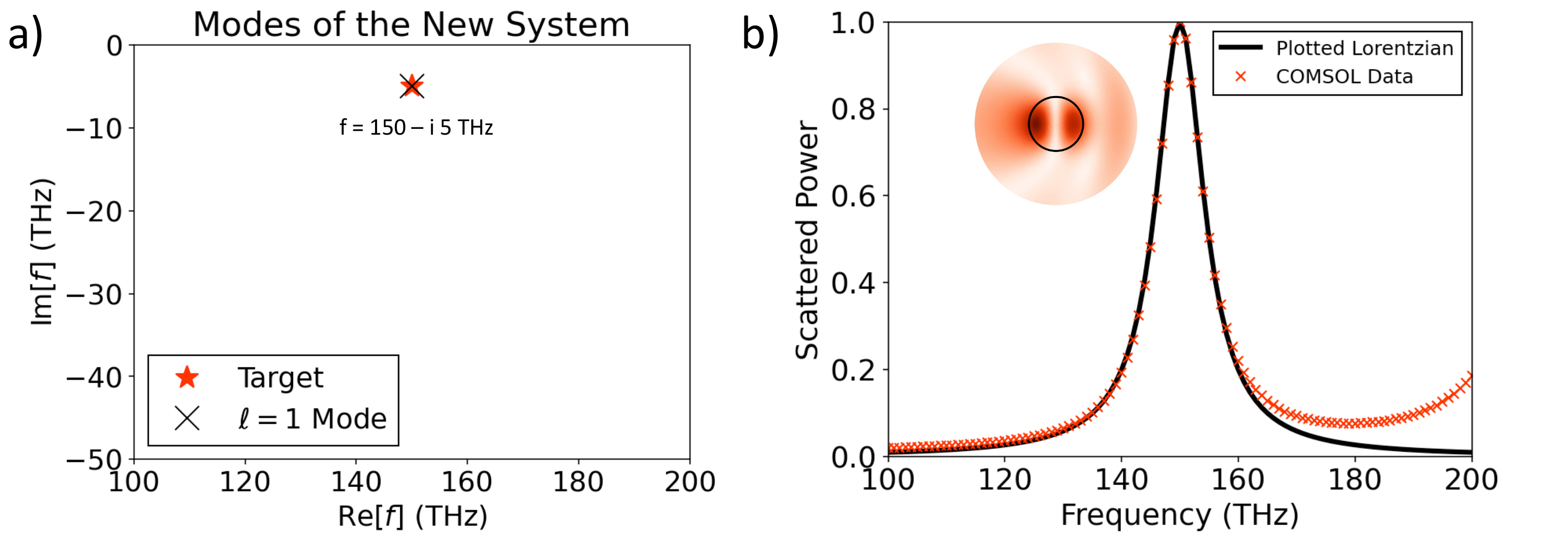}
    \caption{Finding an eigen--permittivity to place the dipole $\ell =1$ mode of an isotropic sphere at $f = 150 - i5$ THz.
    A background permittivity shift of $\varepsilon_b = 0.96 - i0.88$.
    Locations of the modes are shown in a).
    b) Scattered power is then calculated using full--wave simulations, with the mode profile at 150 THz shown inset.}
    \label{fig:sphere_l1}
\end{figure}

We also apply this method to move the $\ell = 1$ dipole mode of a sphere of initial permittivity $\varepsilon = 2$.
Initially, this mode is located at $f = 172 + i 50$ THz.
Setting the target frequency to be $f = 150 - i5$ THz, corresponding to a lower resonance frequency and 10 times smaller linewidth, we solve the eigenvalue problem posed by Eqn. (\ref{eq:ev}) to obtain a permittivity offset of $\varepsilon_b = 0.96 - i0.88$.
The modes of the new system, as well as the scattered power are shown in Figure \ref{fig:sphere_l1}.

\section{Shifting Resonances By Radially Grading the Permittivity}

The next method we propose employs a radial grading of the permittivity to provide control over the spectral location of the quasi--normal modes of a resonator.  
For this design strategy we employ perturbation theory, which for radially symmetric resonators is well developed \cite{ZelDovich1961, Muljarov2010}.
A small change in the permittivity of the the resonator $\delta \varepsilon(r)$ is connected to a change in the position of the quasi--normal mode by the following expression \cite{Leung1994a, Leung1994b}
\begin{equation}
    \delta k_n = -\frac{k_n}{2} \frac{\int_{0}^{a} dr \psi_n^2 (r) \delta \varepsilon (r) }{\int_0^{a} \psi_n^2 \varepsilon (r) dr + \frac{i}{2k_n} \psi_n^2 (a)} ,
\end{equation}
where $a$ is the radius of the resonator and $\psi_n (r)$ is the field profile associated with the mode.
Re-writing the normalisation factor on the denominator as $\langle \psi_n | \psi_n \rangle$ and changing the permittivity at a single point $r_i$ so that $\delta \varepsilon (r) = \Delta \varepsilon \delta (r-r_i)$, we find that the gradient of the location of the quasi--normal mode with respect to the permittivity is 
\begin{equation}
    \pdv{k_n}{\varepsilon} = - \frac{k_n}{2} \frac{ \psi_n^2 (r) }{\langle \psi_n | \psi_n \rangle} .
\end{equation}
The introduction of boundary terms into the norm $\langle \psi_n | \psi_n \rangle$ is necessary as the self--adjointness of the wave operator now depends upon the boundary conditions \cite{Kristensen2020}.
Now, say we would like to place a particular mode at a given complex frequency $k_{\rm target}$.
Defining the figure of merit as 
\begin{equation}
    \mathcal{F} = (k_n - k_{\rm target})^2 ,
    \label{eq:fom}
\end{equation}
we find that the gradient of the figure of merit with respect to the permittivity is 
\begin{align}
    \frac{\partial \mathcal{F}}{\partial \varepsilon} &= (k_n - k_{\rm target}) \frac{\partial k_n}{\partial \varepsilon} \\
    &= -\frac{k_n}{2} (k_n - k_{\rm target}) \frac{\psi_n^2 (r)}{\langle \psi_n | \psi_n \rangle} .
    \label{eq:gradient}
\end{align}
This result is key: we have an analytic expression for how to change the permittivity at every radial position in order to minimise (or maximise) our chosen figure of merit.
Like the adjoint method for designing graded wave--shaping devices \cite{Lalau-Keraly2013, Piggott2015}, this provides a numerically efficient tool for designing graded structures.
In this case, we can design graded cylinders or spheres with 
with specified complex quasi--normal mode frequencies with a chosen multipolar character.

\begin{figure}
    \centering
    \includegraphics[width=\linewidth]{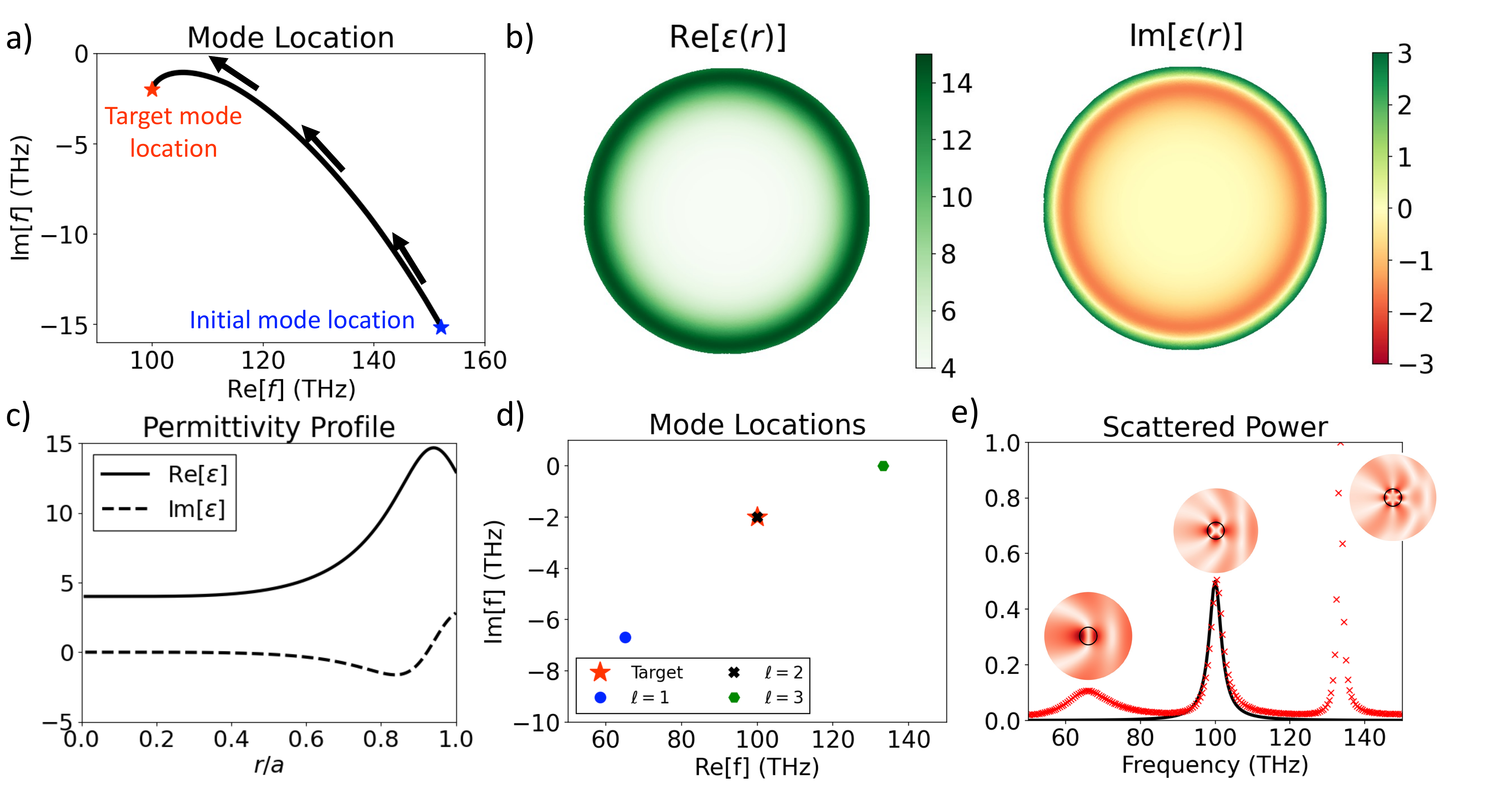}
    \caption{Designing a graded cylinder with a quadrupole resonance at 100-i2 THz.
    a) The iterative movement of the $\ell=2$ mode from 152-i15 THz to the desired complex frequency of 100 - i2 THz.
    Panels b) and c) show the resulting permittivity distribution, which has the pole structure shown in d).
    Full--wave simulations of the scattered power indicate peaks corresponding to each pole, with inset electric fields show that each mode exhibits the expected multipolar nature.}
    \label{fig:graded_cylinder}
\end{figure}

This method is demonstrated in Figure \ref{fig:graded_cylinder}.
Starting from an isotropic cylinder of permittivity $\varepsilon = 4$, the mode locations $k_n$ and fields $\psi_n$ are found using the matrix formulation presented in Section \ref{sec:finding_qnm}.
Choosing to move the $\ell = 2$ mode from $f \sim 250 - i 45$ THz to $f_{\rm target} = 100 - i 2$ THz, the figure of merit we seek to optimise is defined by Eqn. (\ref{eq:fom}).
Evaluating the gradient of the figure of merit according to Eqn. (\ref{eq:gradient}), the permittivity profile is iteratively updated using gradient descent \cite{Shwartz2014} 
\begin{equation}
    \varepsilon^{i+1} (r) = \varepsilon^i (r) + \gamma \frac{\partial \mathcal{F}}{\partial \varepsilon^i} ,
\end{equation}
where the index $i$ indicates the iteration number.
This expression allows for the entire profile to be updated each step.
The progress of the optimisation is shown in Figure \ref{fig:graded_cylinder} a), where the mode moves through the complex plane as the structure is updated.
The designed permittivity grading is shown in Figure \ref{fig:graded_cylinder} b) and c), with the mode locations of the structure shown in panel d).
We note that in order to move the quasi--normal mode towards the real frequency axis, gain is required so that the scattered power is increased.
To verify that the mode has been moved, power scattered from the graded structure has been calculated with finite element full--wave simulations using COMSOL Multiphysics \cite{COMSOL}, with the field distributions associated with each peak shown inset.
A Lorentzian is plotted in Figure \ref{fig:graded_cylinder} e) with a width and central frequency corresponding to the desired location of the quasi--normal mode in the complex plane.

\begin{figure}[h!]
    \centering
    \includegraphics[width=\linewidth]{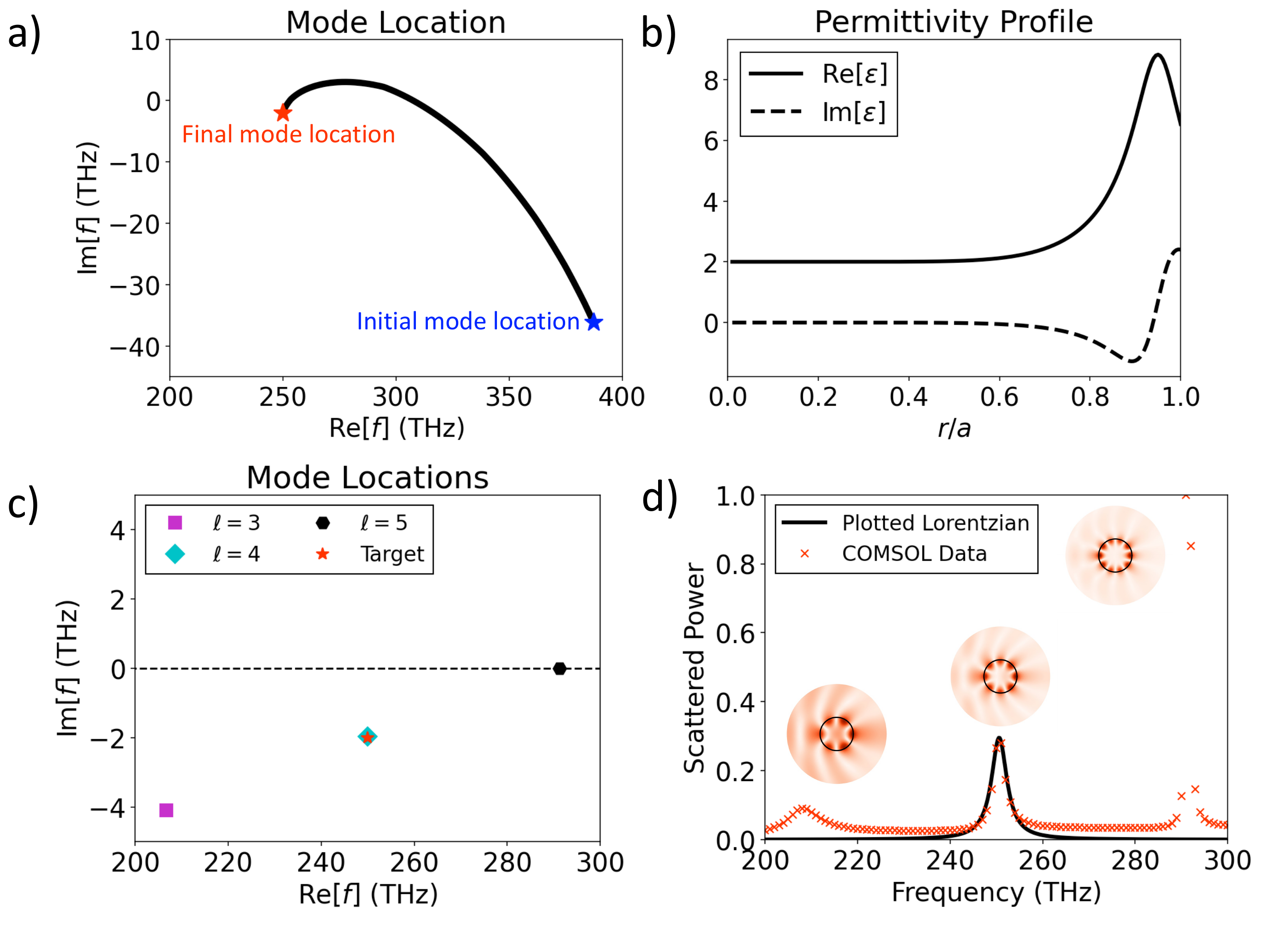}
    \caption{Design of a graded sphere with a $\ell = 4$ resonance at 250 - i5 THz.
    a) Progression of the mode over the iterative optimisation, giving b) the final radial permittivity grading.
    c) Mode locations of the final graded sphere correspond directly to peaks in the d) scattered power.
    Electric field profiles associated with each mode are shown inset.}
    \label{fig:sphere_l4}
\end{figure}
Next, we consider grading the permittivity of an initially isotropic sphere of permittivity $\varepsilon = 2$ to move the $\ell = 4$ resonance to $f_{\rm target} = 250 - i 5$ THz.
Using the same framework, the progression of the optimisation is shown in Figure \ref{fig:sphere_l4} a), with the designed permittivity profile shown in Figure \ref{fig:sphere_l4} b).
The locations of the new modes as well as the scattered power, calculated using full--wave simulations, is shown in Figure \ref{fig:sphere_l4} c) and d) respectively.
In both examples, one can note that most of the grading occurs at distance $r/a > 0.6$.
This is because we are working with modes that vanish at the center of the resonator and manipulating the permittivity in the region where the field is larger gives greater control over the mode location.

\section{Conclusion}

We have developed two techniques to solve the problem of designing cylindrical and spherical resonators with multipolar resonances at desired complex frequencies.
The first approach involves formulating an eigenvalue problem for a complex permittivity shift of the resonator so that a given mode is placed at a particular location.
The second method uses quasi--normal mode perturbation theory to establish a connection between a small change in the permittivity distribution and a small change in the location of a quasi--normal mode.
This is then used to form an analytic expression for the gradient of a figure of merit, taken to be the difference between the location of the mode and a desired location.
With this, the permittivity distribution of the resonator is iteratively updated until the mode is at the desired spectral location.
These methods have the benefit of controlling the resonance frequency of the mode, its linewidth and multipolar nature simultaneously.

We expect our methods to find utility in a range of problems from metamaterial design and sensing to optical computing and communication.
Future developments of our methods might involve including the effects of material dispersion as well as manipulating multiple modes simultaneously.
The ability to arrange the several resonances at desired frequencies would be useful in the design of super--scatterers.
In addition, control over the polarisation of the scattered mode might allow the design of sensors that are polarisation dependent or enable metasurface functionality to be multiplexed.

\begin{backmatter}
\bmsection{Funding}
    We acknowledge financial support from the Engineering and Physical Sciences Research Council (EPSRC) of the United Kingdom, via the EPSRC Centre for Doctoral Training in Metamaterials (Grant No. EP/L015331/1).  
    J.R.C. also wishes to acknowledge financial support from Defence Science Technology Laboratory (DSTL).
    S.A.R.H. acknowledges financial support from the Royal Society (URF\textbackslash R\textbackslash 211033).

\bmsection{Acknowledgments}
    J.R.C. would like to thank J. G. Glasbey for many useful discussions.

\bmsection{Data Availability Statement}
    All data and code created during this research are openly available from the corresponding authors, upon reasonable request.

\bmsection{Supplemental document}
    See Supplement 1 for supporting content. 

\end{backmatter}


\end{document}